# Amark: Automated Marking and Processing Techniques for Ambulatory ECG Data

Sharath Koorathota[1,2], Richard P. Sloan[1]

[1]Columbia University Medical Center, New York NY 10032 USA | [2]Contact: sk4172@columbia.edu

## Abstract

We describe techniques and specifications of MATLAB software to process ambulatory electrocardiogram (ECG) data. Through template-based beat identification and simple pattern recognition models on the intervals between regular heart beats, we filter noisy sections of waveform and ectopic beats. Our end-to-end process can be used towards analysis of ECG and calculation of heart rate variability metrics after beat adjustments, removals and interpolation. Classification and noise detection is assessed on the human-annotated MIT-BIH Arrythmia and Noise Stress Test Databases.

Keywords: electrocardiogram, ambulatory, ectopic, arrythmia, pattern detection

## Introduction

Ambulatory electrocardiogram (ECG) devices are used in the monitoring of patients and research participants in the field, collecting data that may span multiple days. These data can be used to measure heart rate variability (HRV), an index reflecting autonomic modulation of the heart, which has been shown to have clinical significance and can inform researchers about the autonomic effects of stress and activities throughout the day.

HRV analysis requires a time series of R-R intervals, the time between the R peaks for consecutive QRS complexes. Because HRV assesses autonomic control of the heart, the RRIs must represent normal sinus rhythm and be free of artifact and thus, the time series requires detection and correction of non-sinus intervals. Traditionally, this has been accomplished using arrhythmia detectors, which evaluate the QRS morphology from continuous ECG recordings. Increasingly, however, ambulatory monitoring devices generate only RRIs, either because the devices read and interrogate the cardiogram but output only the RRI time series or because the time series comes from a non-ECG source, e.g., a photoplethysmographic (PPG) signal {Jeyhani, 2015 #15947}.

Without the source ECG, artifact identification of the RRI time series is significantly compromised and must rely solely on statistical properties of the RRI series. Some consensus has evolved around a very simple approach: accepting RRIs that fall within a window of acceptable length (e.g., greater than 0.3 seconds and less than 1.8 seconds). Nevertheless, recent approaches to analysis of patterns of RRIs, not just their length, have been employed successfully to accomplish the detection and correction with high accuracy and efficiency in practice [1]. These methods control for variations caused by exercise or other types of motion when calculating HRV measures[2], [3].

Regardless of whether the RRI time series derives from a continuous ECG recording or comes from a non-ECG based device, the identified artifacts must be corrected, if possible, before submitting the data for analysis of HRV. The most commonly implemented approach is to delete the artifactual RRI and replace it by linear interpolation. Caution in using this approach is required because excessive interpolation by definition reduces variability. There are no generally accepted standards for what constitutes an acceptable amount of interpolation. On the other hand, uncorrected artifact can significantly distort HRV analysis as Figure XX clearly demonstrates.

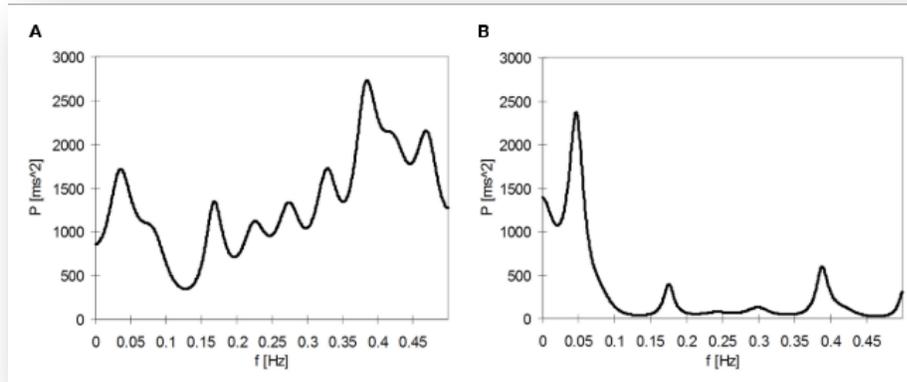

*Figure 1: (A) Power spectrum of a 3-min segment of the RRI time series containing a ventricular premature beat. (B) Corresponding power spectrum, where the VPB is edited. Adapted from Peltola (2012).*

Smoothing RRI series from ambulatory recordings to correct artifacts and reflect the underlying HRV dynamics has been discussed extensively [4][5][6], but the specific criteria for determining which irregularities should be corrected and which should be excluded from HRV analysis are underreported.

With ECG data, regions of the RRI time series may be classified as irregular and be excluded from analysis if the surrounding signal contains noise or through information contained in the cardiogram (e.g. absence of an accompanying P-wave). However, the waveform data are not always available, e.g., when RRIs are derived from non-ECG methods like photoplethysmography or when a waveform recording contains electrical noise that obscures the R-peak.

We propose a set of algorithms (packaged in a freely-available software labeled "Amark") that process RRI series prior to HRV or other analysis. Below, we primarily discuss the outlier detection and correction methods in cases with an ECG data source, but similar processing methods may be extended to non-ECG sources.

## Computational Methods

Amark processes ECG data from a single ECG channel in three stages (Figure 2). The first, beat identification, identifies R peaks in the ECG waveform through template matching and peak-detection methods. The second, irregularity detection, uses regional noise profiles, RRI patterns and P-wave detection to identify missed beats, noise and arrhythmia. The third, correction, attempts to classify a region of the RRI time series as usable or unusable for HRV analysis (i.e. unable to be corrected) by smoothing certain RRI tachogram patterns in irregular regions.

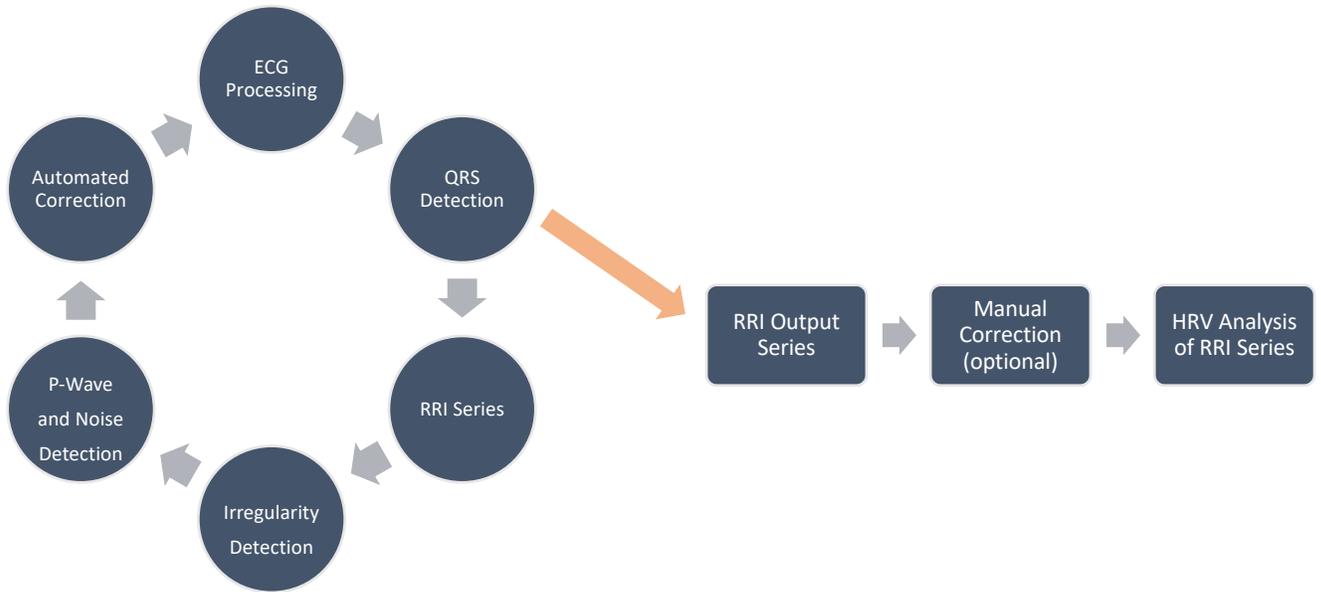

*Figure 2: The beat identification stages of Amark include ECG processing and QRS detection (using the GQRS algorithm), while the irregularity detection uses both patterns in RRI series, and noise quality and P-wave around each QRS complex. After the irregularity detection, Amark attempts to correct through automated methods. The signal may be processed again and the output RRI series are used in HRV analysis. Any irregularities at this stage can be manually corrected by the user prior to HRV analysis.*

## Beat Identification

For beat identification, Amark uses multiple, open-source algorithms available as part of the PhysioToolkit [7]. The novel gqrs and gqpost detectors locate QRS complexes with high sensitivity and post-process the annotations to improve predictivity.

## Irregularity Detection

The goal of irregularity detection is not to classify all possible types of arrhythmias but to maximize the amount of usable data in ambulatory recordings, identify infrequent arrhythmias that lead to irregularities in the RRI series, and to correct those irregularities prior to calculating HRV measures.

Amark uses layers of decision rules on the RRI series and underlying ECG to identify arrhythmia or annotation errors resulting from the beat identification stage, based on a simplified implementation of algorithms first proposed by Citi et al [5], which evaluate a probability density of observing a beat given a previous beat. In Amark, classifications rely on layers of decision rules to classify beats as normal or irregular, similar to other, previously-published methods [8]. The rules make use of ecgpuwave [9], an independent QRS detector and waveform limit locator used to detect the presence of P-waves, and a noise detector to distinguish between valid sinus rhythm and irregularities.

An irregularity could belong to one of several categories and may occur in single beats, pairs or multiple consecutive beats (Table 1).

| Identification related | Arrhythmia related |
| --- | --- |
| Extra beat | Paced beat |
| Missed beat | Atrial premature beat |
| Misplaced beat | Ventricular escape beat |
| Signal quality related | Fusion of paced and normal beat |
| Noisy beat | Nodal (junctional) premature beat |
| Ventricular flutter waves | Left bundle branch block beat |
| End of ventricular flutter/fibrillation | Unclassifiable beat |
| Isolated QRS-like artifact | Right bundle branch block beat |
| Change in signal quality | Supraventricular premature or ectopic beat (atrial or nodal) |
| Rhythm change | Premature ventricular contraction |
| Non-conducted P-wave (blocked APC) | Unidentified complexes |

*Table 1: Summary of irregularities targeted by Amark. These are extensively catalogued in the MIT-BIH Arrythmia Database.*

### Correction

RRI series may be corrected in three possible ways:

1. Removing extra beats,
2. Adding beats (interpolation),
3. Adjusting the timing of beats.

Irregularities too extreme to correct are labeled to be excluded from subsequent analysis.

## Software description

Amark 1.0 has been developed using MATLAB® release 2018b (The MathWorks, Inc.) and was compiled to a deployable standalone application with the MATLAB® Compiler 9.0. The MATLAB® Compiler Runtime (MCR) version 9.3 is required for running Amark.

### Data formats

Amark reads single-lead ECG data. The sample rate is specified in the file itself or in the program window. Input data could be one of the options below (at least 2 minutes of data are required), and all output data formats are exported after analysis.

| Input Data Format | Description | Output Data Format | Description |
|---|---|---|---|
| .txt | Tab-delimited | .rtimes | Times of valid QRS complexes (post-correction) |
| .edf | European data format | .bi | "Bad" or irregular intervals |
| .bdf | Modified EDF files | _output.mat | File |
| .mat | MATLAB files with signal amplitudes and optional frequency variables | | |
| _output.mat | Previously-processed Amark files used for viewing results and parameters | | |

*Table 2: Data formats for input and output of Amark.*

## The Primary User Interface (PUI)

Amark's PUI contains three sections: (1) data upload and initialization, (2) parametrization, and (3) processing.

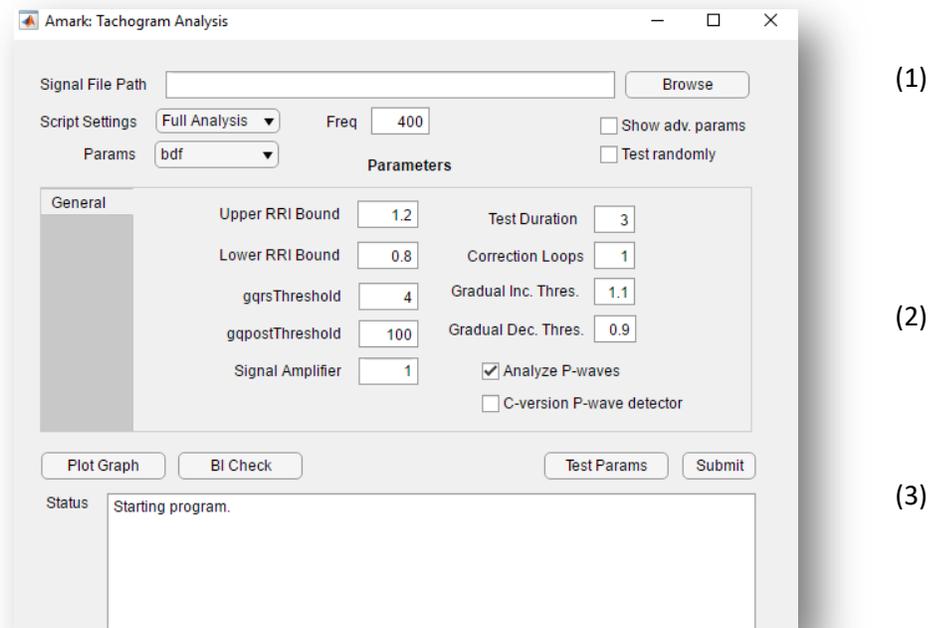

*Figure 3: Amark's primary user interface.*

(1) Data upload and initialization

Raw ECG, HR or RRI series data is initialized after upload through the interface. The sampling frequency can be specified along with preset parameters to use. There also is an option to select a random region while testing effectiveness of parameters, or to re-use the same testing region to compare performances between different sets of parameters.

(2) Parametrization

This parametrization panel displays the primary parameters used most often with Amark (detailed in the "Software Parameters" section below).

(3) Processing

The processing panel is used to view the raw or processed ECG waveform ("Plot Graph"), check the post-processing, irregularity regions, process a short, test region of the input file with the given parameters ("Test Params"), or submit the full, input ECG file for processing ("Submit").

Software Parameters

The standard set of parameters in Amark spans three categories: those that relate to the beat detection stage (1-3 in Table 3), regional thresholds (4-7), parameter testing (8), and correction-related (9-11).

| Reference No. | Primary Parameters | Description |
| --- | --- | --- |
| 1 | gqrsThreshold | Threshold for Physionet's gqrs detectors |
| 2 | gqpostThreshold | Threshold for Physionet's gqpost detectors |
| 3 | Signal Amplifier | Amplification constant for signal |
| 4 | Upper RRI Bound | Inner, upper range for outlier detection – any RRI value above this regional threshold will be excluded and conditional checks (ex. for valid sinus rhythm patterns) are run on intervals that fall outside of the threshold. |
| 5 | Lower RRI Bound | Inner, lower range for outlier detection – any RRI value below this regional threshold will be excluded and conditional checks (ex. for valid sinus rhythm patterns) are run on intervals that fall outside of the threshold. |
| 6 | Gradual Inc. Thres. | Sensitivity to gradually increasing RRI regions, similar to upper RRI bound but with a shorter (1 beat) window. |
| 7 | Gradual Dec. Thres. | Sensitivity to gradually decreasing RRI regions, similar to upper RRI bound but with a shorter (1 beat) window. |
| 8 | Test Duration | Amount of time (sec) on which to run the analysis for testing purposes. The tested segment can be randomly selected or a previously-tested region may be reused. |
| 9 | Correction Loops | Number of times to loop through the process for outlier detection (identification, correction, analysis, p-wave detection). Correcting marks changes regional statistics, so more marks may be corrected in second passes. The downside of running more loops is overcorrection of the data. |
| 10 | Analyze P-waves | Implements the MATLAB version of the ecgpuwave delineator |
| 11 | C-version P-wave detector | Implements a C-version the ecgpuwave algorithm, much more sensitive and prone to errors |

Table 3: The main parameters used in Amark.(1-3) deal with the beat-detection stage, (4-7) control regional thresholds, (8) deals with parameter testing, (9-11) are related to irregularity correction.

## Data Viewer

The raw ECG waveform or post-processed results of the input file or test region are viewed in the "Data Viewer" window that is automatically displayed after processing or opened manually from the PUI.

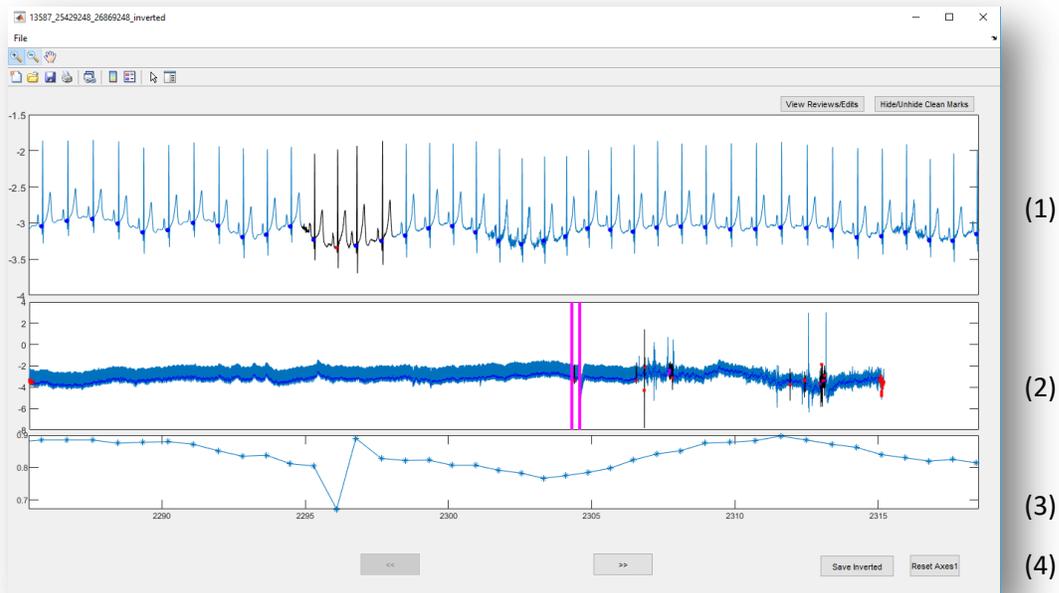

*Figure 4: Viewer window with (1) sub-region panel, (2) full waveform, (3) tachogram.*

The processing results are displayed in three panels. The sub-region panel (1) displays the zoomed in region of the processed waveform, with markings generated by Amark. This is a helpful step to assess whether the corrections appear valid. Table 4 contains a color glossary of each beat type with description. After the identification stage but before correction, the peaks can be reviewed by clicking the "View Review/Edits" button above the sub-region panel. This option allows the user to review the types of edits Amark performed (if any) during the correction stage. The full waveform (2) is displayed with the sub-region delineated with vertical purple line markers. The tachogram panel (3) displays the length (in seconds) of RR intervals of the marks shown in the sub-region panel. Amark also allows the viewing of raw ECG waveforms through the Viewer and, to correct cases where electrodes have been placed incorrectly on a participant, the ability to invert the files through the bottom panel (4).

| Symbol | Description | Example |
|---|---|---|
| Blue Waveform | Valid sinus rhythm or corrected regions | 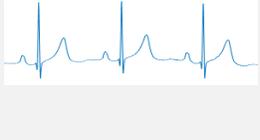 |
| Black Waveform | Non-sinus beats or uncorrectable regions | 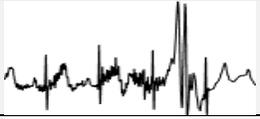 |
| Blue Dot | Valid QRS complexes or corrected peaks | 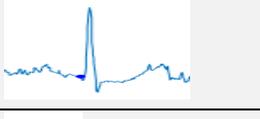 |
| Red Dot | Invalid QRS complexes or peaks associated with uncorrectable RRI regions. | 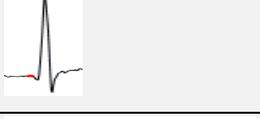 |
| Green Dot | QRS complexes corresponding to short intervals which can be removed for smoother tachogram (possibly an extra beat added due to noise) | 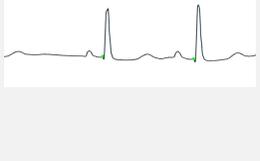 |
| Pink Dot | Beat pairs of a short-long interval pattern. Typically PVC beats. | 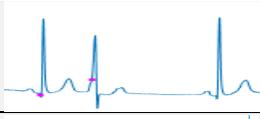 |
| Cyan Dot | Beats with associated, atypically long RRIs that are able to be interpolated. | 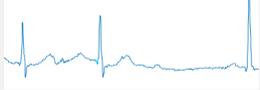 |

*Table 4: Marks after identification and classification stages.*

Irregularity Review

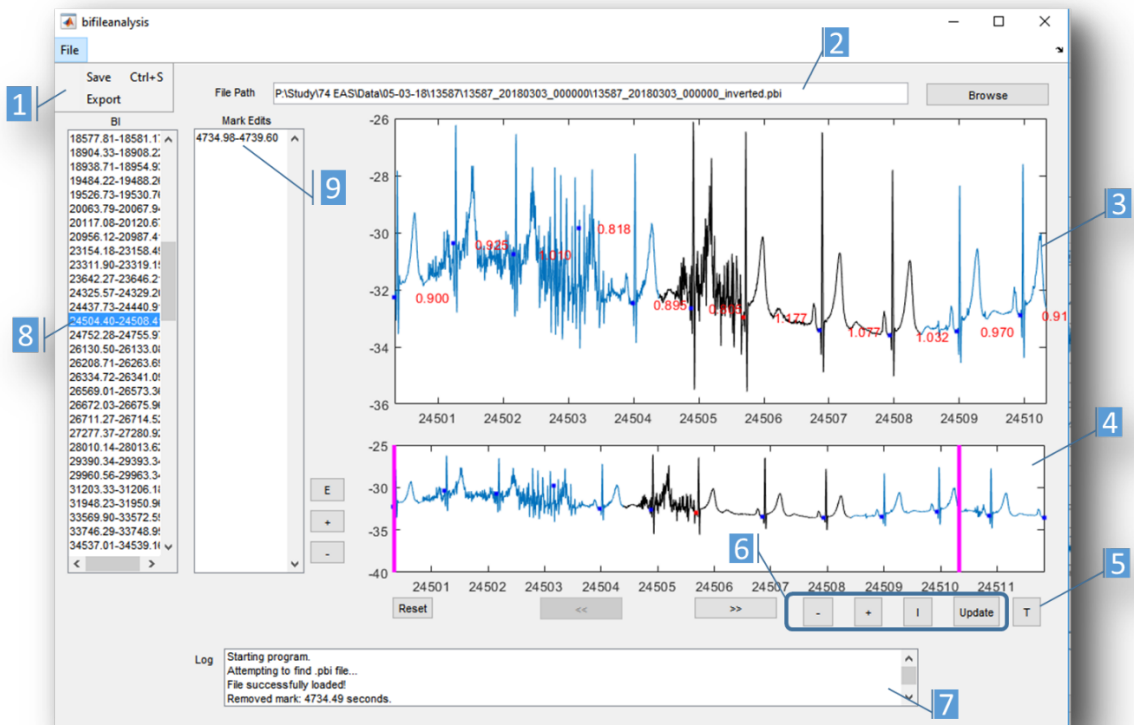

*Figure 5: The Bad Interval window, showing regions of the ECG waveform with irregularities that were not corrected.*

The results of processing also can be viewed through the Irregularity Review window. This window is used to view any identified intervals from the ECG waveform that contain irregularities or arrhythmias that could not be corrected using the specified parameters. Processed files are uploaded (section 2 in Figure 5) and the regions containing bad intervals are identified in (8), where the user can modify them. Any manual modifications are recorded in the log (7) and recorded in the "Mark Edits" column (9). Selected regions are displayed in the viewing panel (3), where the exact region marked as an irregular interval appears in black and the surrounding, artifact-free, region appears in blue for context. The RRI duration is displayed in red font and the colored points indicate the onset of the QRS complexes. The vertical, purple lines displayed in a separate panel (4) identify the viewing window.

The Irregularity Review window also allows for relocation of the marks identifying R peaks, and thus the RRIs (section 5). Marks can be deleted (-), added (+), or interpolated (I) after selecting a beat and specifying the number of additional beats to linearly interpolate in the associated interval.

# Correction Algorithms
Amark corrects beats initially classified as "excluded" (most likely arrhythmia- or noise-induced) through removing, adding or adjusting beats.

### Noise Detection

A noise profile around each detected beat is calculated through the following steps:

1. The derivative of the ECG signal is calculated at each sample within a window (e.g., 200 ms) before and after each beat,
2. The variance of this signal derivative is averaged over the window, resulting in a "noise profile,"

This simple measure has been validated through testing and performs efficiently for ambulatory data (Table 6).

### Outlier Detection

After the initial beat identification stage, outliers are detected using the RRI series. The average of the 20 RRI durations preceding and following each beat, in addition to their associated noise profiles, are used to detect outlier beats whose RRIs or noise profiles are below 80% or above 120% of the regional averages (referred to as the regional threshold, see reference #4 and #5 in Table 3). Next, Amark calculates the cumulative sums of the 4 RRIs after each beat. This metric is used to find the number of adjacent beats whose RRI is closest to the regional average. Pairs of adjacent outlier beats are assessed to determine if their RRI sum is approximately double the regional average, within the regional threshold. This pattern is typical of several types of premature beats. Next, long, outlier RRIs are split into even-duration sub-intervals, which are assessed to determine if their durations are within the regional threshold. If the outlier RRIs can be validly split, the beats associated with the RRIs are marked as able to be interpolated later in the correction process.

### P-wave Based Classifications

To assist the classification of outlier beats, Amark implements the ecgpuwave algorithm (see Beat Identification section). If a P-wave is not found for an outlier beat ($B_n$), Amark checks if it is isolated (i.e. $B_{n-1}$ and $B_{n+1}$ are valid). The program reviews all such isolated beats ($B_i$) to assess if they have a smaller RRI than the preceding beats ($B_{i-1}$) and classify all $B_i$ as the longer beats of short-long, irregular patterns characteristic of premature beats. All beats classified as $B_i$ are corrected by adjusting the time of $B_i$ so that RRIs associated with $B_{i-1}$ and $B_i$ are of equal length (beat type BT1 in Table 5). This essentially treats the initial annotations as misplaced, and corrects them to yield a smoother tachogram [5].

If a P-wave for an irregular beat, $B_y$, is found, Amark first checks if the next, adjacent beat, $B_{y+1}$, also is irregular. In this case, the program excludes both $B_y$ and $B_{y+1}$ since the waveform surrounding the beats most likely contains the type of non-sinus patterns typical of premature atrial contractions (BT2 in Table 5). If $B_y$ is isolated ($B_{y-1}$ and $B_{y+1}$ are included), Amark checks if the change in RRIs associated with $B_{y-1}$ and $B_y$ is gradual, or falls within a specified, regional threshold (see reference #6 in Table 3). If so, $B_y$ is included (BT3 in Table 5). If the RRI associated with $B_y$ is higher than the physiological, upper bound (> 1.5 seconds), then $B_y$ is classified as irregular (BT4 in Table 5). Similarly, the program checks RRIs for multiple, adjacent, excluded beats for gradual shifts in the tachogram.

### Corrections

After modifications of beats in the P-wave classification stage, Amark re-analyzes irregular beats. First, the program removes remaining extra beats using the cumulative sums from the outlier detection stage. Beats in the sequence are removed so that the resulting tachogram is smoothed, which may require the removal of multiple, adjacent beats (BT6 in Table 5).

Then Amark interpolates outlier RRIs, of eligible length, to yield multiple sub-intervals of equal length, closer to the regional average (BT7 in Table 5). Regional, outlier detection measures are re-calculated after interpolation. Lastly, remaining short-long RRI sequences are smoothed by averaging the two RRIs and adjusting the time of the second beat so that the resulting RRIs are closer to the regional average (BT8 in Table 5). For example, if the regional average RRI is 0.8 seconds at $B_a$, but the RRI at $B_a$ is 0.6 seconds and the RRI at $B_{a+1}$ is 1.0 seconds, Amark adjusts the timing of $B_{a+1}$ by 0.2 seconds so that both $B_a$ and $B_{a+1}$ have a resulting RRI that is 0.8, or the regional mean.

The number of P-wave classification and correction periods are controlled through the "Correction Loops" parameter (Table 3) and outlier detection is completed on the full RRI series after each loop.

| Beat type | Description | Stage | Result |
|---|---|---|---|
| BT1 | Short-long (SL) pairs typical of PVC patterns (P-wave not found) | P-wave | Adjusted |
| BT2 | Typical PAC patterns (P-wave found) | P-wave | Excluded |
| BT3 | Gradual increases (P-wave found) | P-wave | Included |
| BT4 | Sudden increases (P-wave found) | P-wave | Excluded |
| BT5 | Gradual decrease (P-wave found) | P-wave | Included |
| BT6 | Extra beats (typically noise) | Correction | Removed |
| BT7 | Beats with large RRIs, close to regional average | Correction | Interpolated |
| BT8 | SL pairs typical of PVC patterns | Correction | Adjusted |

*Table 5: Corrections for typical beats. Beats that are "Excluded" are left as too irregular to fix. The initial, irregular classification is removed for" Included" beats. "Adjusted" beats have the time of the QRS complex adjusted so to yield a smoother tachogram.*

### Irregularity Regions

The first and last 20 beats of the record are used to train the algorithm by calculating regional average noise quantity and RRIs that can inform the approach to the rest of the file. Thus, Amark always marks these 20-beat training sections to be excluded from the resultant RRI time series. Next, for other excluded, irregular beats ($B_e$), the program starts the bad interval at the midpoint between the last valid beat, $B_{e-1}$, and the irregular one, $B_e$. Amark ends the irregular region at the midpoint between the last irregular beat, $B_e$, and the next valid one, $B_{e+1}$. Amark continues this process until reaching the next valid beat or reaching the end of the file.

## Validation

We assessed the performance of our correction algorithms on the MIT-BIH database [7] and the 20 annotation categories present in 48 half-hour excerpts of single-channel ambulatory ECG recordings (Figure 6). Amark classifies approximately 87.7% of beats as "valid" after the identification and correction stages with 6.7% of marks present in the database not identified post-correction (labeled as "Not Present"). On average, the correction algorithms increase the percentage of valid data by 9.8%, across all classifications of beats. This represents an improvement in the quality and quantity of usable data, through correction of non-sinus rhythm through beat adjustment and interpolation.

| Classifications | Post-Correction Counts | | | Change after Identification Proportions | | |
|---|---|---|---|---|---|---|
| Description | Irregular | Valid | Not Present | Irregular | Valid | Not Present |
| Ventricular flutter wave | 27 | 126 | 319 | -0.1292373 | 0.116525 | 0.012711864 |
| Paced beat | 192 | 6819 | 17 | -0.0032726 | 0.00683 | -0.0035572 |
| Start of ventricular flutter/fibrillation | | | 6 | 0 | 0 | 0 |
| End of ventricular flutter/fibrillation | | 3 | 3 | 0 | 0.5 | -0.5 |
| Isolated QRS-like artifact | 5 | 19 | 108 | -0.0151515 | 0.045455 | -0.03030303 |
| Change in signal quality | 1 | 17 | 598 | -0.0016234 | 0.021104 | -0.019480519 |
| Rhythm change | | 37 | 1253 | 0 | 0.027907 | -0.027906977 |
| Atrial premature beat | 254 | 2028 | 414 | -0.4187685 | 0.309347 | 0.109421365 |
| Ventricular escape beat | 8 | 113 | 1 | -0.0409836 | 0.04918 | -0.008196721 |
| Fusion of paced and normal beat | 85 | 1686 | 14 | -0.0207283 | 0.02465 | -0.003921569 |
| Nodal (junctional) premature beat | 73 | 213 | 26 | -0.3557692 | 0.272436 | 0.083333333 |
| Left bundle branch block beat | 140 | 7875 | 60 | -0.0185759 | 0.019071 | -0.000495356 |
| Normal beat | 4454 | 68819 | 1779 | -0.0445957 | 0.02926 | 0.015336034 |
| Unclassifiable beat | 1 | 26 | 6 | -0.0909091 | 0.121212 | -0.03030303 |
| Right bundle branch block beat | 388 | 6761 | 110 | -0.0512467 | 0.037333 | 0.013913762 |
| Supraventricular premature or ectopic beat (atrial or nodal) | | 2 | | 0 | 0 | 0 |
| Premature ventricular contraction | 776 | 3801 | 2557 | -0.3451586 | 0.027402 | 0.317756938 |
| Non-conducted P-wave (blocked APC) | | 10 | 183 | 0 | 0.051813 | -0.051813472 |
| Not Identified by Physionet | 164 | 4620 | | -0.2018302 | 0.20183 | 0 |
| Comment Annotations | | 6 | 431 | **Average Change** | | |
| Total | 6568 | 102981 | 7885 | | | |
| Proportions | 0.055929288 | 0.876926614 | 0.067144098 | -0.0914658 | 0.097966 | -0.006500241 |

*Figure 6: A summary of classifications in the MIT-BIH arrhythmia database. Most of the initially excluded marks (during the identification stage) are corrected by Amark to result in a smoother tachogram. In some cases, Amark identified marks that were not present on Physionet, labeled as "Not Identified by Physionet." In other cases, Amark did not find beats that were labeled by Physionet, labeled as "Not Present."*

We also independently report the performance of our noise detection algorithm on 12 half-hour ECG recordings reported with varying levels of signal-to-noise ratios from the MIT-BIH Noise Stress Test

Database [7]. Overall, for SNR less than or equal to 12 db, the accuracy of the noise classifications (i.e. correctly assessing whether a region contained noise) was approximately 83.7%.

| Identifier | SNR (db) | Accuracy |
|---|---|---|
| 118e24 | 24 | 0.514 |
| 118e18 | 18 | 0.542 |
| 118e12 | 12 | 0.711 |
| 118e06 | 6 | 0.824 |
| 118e00 | 0 | 0.879 |
| 118e_6 | -6 | 0.859 |
| 118e24 | 24 | 0.599 |
| 118e18 | 18 | 0.641 |
| 118e12 | 12 | 0.779 |
| 118e06 | 6 | 0.88 |
| 118e00 | 0 | 0.899 |
| 119e_6 | -6 | 0.867 |

Table 6: Performance of the simple noise detection algorithm used in Amark. The average accuracy in noise classification is 83.7% for SNR < 12.

Because the MIT-BIH arrhythmia database does not suggest how specific arrhythmias should be treated for HRV analysis, we report summary measures, including the number of epochs that were analyzed by an in-house spectral algorithm (Table 7). Amark accurately identified 93.1% of beats, with a precision rate of 97.3%, post-correction. The spectral algorithm we utilized searches for 5-min epochs with RRIs typical of regular sinus rhythm. Amark, through beat adjustments and corrections, increased the number of usable epochs in the MIT-BIH database by 55.9% on average across files.

| Measure | Description | Post-Identification | Post-Correction | MIT-BIH |
|---|---|---|---|---|
| Accuracy | Synced Beats/Total Beats on MIT | 0.962994692 | 0.930677558 | - |
| Precision | Total Amark Beats/Total Beats on MIT | 0.965880164 | 0.973460313 | - |
| Proportion Normal | Included on Amark & 'N' on MIT/Total 'N' on MIT | 0.870222035 | 0.898243648 | - |
| PVC Proportion | PVC Found/Total PVC on MIT | 0.891391167 | 0.567028825 | - |
| PAC Proportion | PAC Found/Total PAC on MIT | 0.964169193 | 0.805499165 | - |
| PVC Included Proportion | PVCs Included/Total PVC on MIT | 0.584236529 | 0.475487306 | - |
| PAC Excluded Proportion | PACs Excluded/Total PAC on MIT | 0.387395456 | 0.141709991 | - |
| Spectral Epochs | Total Spectral Epochs | - | 92 | 59 |

Table 7: Proportions of beats, post-identification and correction stages, along with overall accuracy and precision measures. We also include the proportion of beats for two arrhythmias of special interest, premature ventricular contractions (PVC) and premature atrial contractions (PAC) and the number of epochs that were processed through an in-house spectral algorithm.

## Availability of software

Amark is available as a MATLAB software package upon request. Please contact the author.